\documentstyle[11pt]{article}
\begin{document}
\baselineskip 14pt
\centerline{\Large\bf Local U(1) symmetry in Y(SO(5)) associated }

\centerline{\Large\bf with Massless Thirring Model and its Bethe Ansatz }
\vspace{3cm}
\centerline{ Hong-Biao Zhang$^{1,2}$ ,  Mo-Lin Ge$^1$, Kang Xue$^{1,3}$ }
\vspace{0.5cm}
{\small
\centerline{\bf 1. Theoretical Physics Division, Nankai Institute of Mathematics,}
\centerline{\bf Nankai University, Tianjin 300071, P.R.China}
\centerline{\bf 2. Educational Institute of Jilin Province, }
\centerline{\bf Changchun,Jilin,130022, P.R.China}
\centerline{\bf 3. Physics Department, Northeast Normal University,}
\centerline{\bf Changchun,Jilin, 130024, P.R.China}
}
\vspace{2cm}

\centerline{\bf Abstract}


\vspace{3cm}

\pagebreak


{\bf (I) Introduction}

    Recently, it has been proposed by S.C.Zhang et al. that the
 antiferromagnetic(AF) and superconducting(SC) phases of high-$T_c$
 cuprates are unified by an approximated $SO(5)$ symmetry principle[1].
 Considerable support for this proposal came from numerical investigations
 in models for high-$T_c$ materials. In particular, it was shown that the
 low-energy excitations can be classified in terms of an $SO(5)$ symmetry
 multiplet structure[2,3]. Subsequently, extended Hubbard models and a
 two-leg ladder model related to $SO(5)$ symmetry have been introduced
 and analyzed in details[4,5,6]. On the other hand, Shelton and
 S$\acute{e}$n$\acute{e}$chal[7] have studied the problem of two coupled
 1D Tomonaga-Luttinger chains and concluded that approximate $SO(5)$
 symmetry can emerge in low-energy limit of this model. It is well-known
 that the Luttinger liquid is connected with the massless Thirring model.
 It is worth to deal with massless Thirring model with $SO(5)$ symmetry.
 The model can be constructed by four-component fermions field operator
 $\psi_{i}(x)$, we shall show that this model is exactly solvable
by the Bethe ansatz method through a local $U(1)$ transformation, under which
the fermion operator $\psi_{i}(x)$ is transformed into q-deformed fermion
operator $\Phi_{i}(x)$. This procedure leads to the diagonalization that looks
in a simple manner by Wadati[8,9,10]. Furthermore, the study of Yangian
algebra[11,12,13,14] provides a significant tool in the formalism of integrable
models. The generators of Yangian can be realized through currents for a given Lie algebra.
It turns out that the current realization of $Y(SO(5))$ is not unique and
allow a local $U(1)$ gauge transformation. It is interesting to find the
consequence of such a $U(1)$-freedom according to the q-deformation of
fermion operator $\Phi_{i}(x)$.

     This paper is organized as follows: In the section (II),
  the massless Thirring model with $SO(5)$ symmetry will be diagonalized
and the Bethe ansatz wavefunction is constructed. In section (III), we shall
give the current algebra realization of $Y(SO(5))$ in terms of q-deformed
fermion current that give rise to the local $U(1)$-gauge transformation.

{\bf (II) The massless Thirring model with $SO(5)$ symmetry and its Bethe ansatz wavefuction}

        Let us consider the massless Thirring model constructed by the
 four-component fermion field operator $\psi(x)=[\psi_{1}(x),\psi_{2}(x),\psi_{3}(x),\psi_{4}(x)]^T$.
The Hamiltonian takes the form:
\begin{equation}
\label{e1}
H=\int [ iv\sum_{i=1}^{4}C_i \psi_i ^+(x) \partial_x \psi_i(x)+
g\sum_{i,j=1}^{4} C_{ij}n_i(x)n_j(x)] dx
\end{equation}
where $C_{ij}=C_{ji},C_{ii}=0$ and $n_i(x)=\psi_i^+(x) \psi_i (x)$
 ($i,j=1,2,3,4$) satisfy the anticommutation relations:
\begin{equation}
\label{e2}
  [{\psi_i^+ (x)},{\psi_j^+ (y)}]_+=0
\end{equation}
\begin{equation}
\label{e3}
 [{\psi_i (x),{\psi_j (y)}}]_+=0
 \end{equation}
 \begin{equation}
 \label{e4}
 [{\psi_i (x)},\psi_j^+ (y)]_+=\delta _{ij} \delta (x-y)
\end{equation}

  For four-component fermion field operators $\psi(x)=[c_\sigma(x),d^+_\sigma(x)]^T$
and forms the current algebra obeying $SO(5)$[6]. In momentum space,
this Hamiltonian can be written as:
$$
H=\int [ -v\sum_{i=1}^{4}kC_i n_i(k)] dk
$$
$$
+\frac{g}{\pi}\int\int\int[C_{12}c_\uparrow^+(k+\frac{q}{2})
c_\downarrow^+(-k+\frac{q}{2})c_\downarrow (-k'+\frac{q}{2})
c_\uparrow (k'+\frac{q}{2})
$$
$$
+C_{13}c_\uparrow^+(k+\frac{q}{2})
d_\uparrow^+(-k+\frac{q}{2})d_\uparrow (-k'+\frac{q}{2})
c_\uparrow (k'+\frac{q}{2})
$$
$$
+C_{14}c_\uparrow^+(k+\frac{q}{2})d_\downarrow^+(-k+\frac{q}{2})
d_\downarrow (-k'+\frac{q}{2})c_\uparrow (k'+\frac{q}{2})
$$
$$
+C_{23}c_\downarrow^+(k+\frac{q}{2})d_\uparrow^+(-k+\frac{q}{2})
d_\uparrow (-k'+\frac{q}{2})c_\downarrow (k'+\frac{q}{2})
$$
$$
+C_{24}c_\downarrow^+(k+\frac{q}{2})d_\downarrow^+(-k+\frac{q}{2})
d_\downarrow (-k'+\frac{q}{2})c_\downarrow (k'+\frac{q}{2})
$$
\begin{equation}
\label{e5}
+C_{34}d_\uparrow^+(k+\frac{q}{2})d_\downarrow^+(-k+\frac{q}{2})
d_\downarrow (-k'+\frac{q}{2})d_\uparrow (k'+\frac{q}{2})] dk dk'dq
\end{equation}
that obviously is made up of the pairs, so  this model may be applied to superconducting.

      To diagonalize H, we introduce local $U(1)$ transformation:
\begin{equation}
\label{e6}
\Phi_i (x)=exp[-i\sum_{k=1}^{4}\theta_{ik}\phi_k (x)] \psi_i (x)
\end{equation}
where $\phi_i (x)=\int_{-\infty}^{x} \psi_i^+ (y) \psi_i (y) dy$ and $\theta_{ik}$ are constants.

    According to eq.(\ref{e2})-eq.(\ref{e4}) and eq.(\ref{e6}) by direct calculation, we
 obtain(no summation over the repeated $j$):
 \begin{equation}
 \label{e7}
\Phi_i (x) \Phi_j (y) =-exp[i\theta_{ij}] \Phi_j (y) \Phi_i (x)
\end{equation}
\begin{equation}
\label{e8}
\Phi_i^+ (x) \Phi_j^+ (y) =-exp[i\theta_{ij}]\Phi_j^+(y) \Phi_i ^+(x)
\end{equation}
\begin{equation}
\label{e9}
\Phi_i (x)\Phi_j^+ (y) =-exp[-i\theta_{ij}]
\Phi_j^+ (y)\Phi_i (x) + \delta _{ij} \delta (x-y)
\end{equation}
   This is a special case of Zamolodchikov-Faddeev algebra[15,16].
\begin{equation}
\label{e10}
\theta_{ii}=0 \;\;\;\ (mod \;\ 2\pi)
\end{equation}
\begin{equation}
\label{e11}
\theta_{ij}+\theta_{ji}=0 \;\;\;\ (mod \;\ 2\pi)
\end{equation}
Therefore, eq.(\ref{e10}) and eq.(\ref{e11}) are conditions given by the associativity
of the special case of Zamolodchikov-Faddeev algebra. The meaning of eq.(\ref{e10})
is clear that the particle itself must still be fermion for the same "i-spin"
states, however, eq.(\ref{e11}) show that the commutation relations between different
"i-spin" states can be q-deformed and the q-deformation parameters should obey
eq.(\ref{e11}) because of two-body interaction between different "i-spin" states.

     Under the local $U(1)$ transformation eq.(\ref{e6}), the Hamiltonian eq.(\ref{e1}) can
be diagonalized and find the physical constraint conditions for real $C_i$ and
$C_{ij}$. The Heisenberg equation $i\partial_t{\psi}_i(x,t)=[\psi_i(x,t),H]$ reads:
\begin{equation}
\label{e12}
 \partial_t {\psi}_i(x,t)=vC_i \partial_x \psi_i(x,t)-
                  i2g\sum_{j=1}^{4} C_{ij}n_j(x,t)\psi_i(x,t)
\end{equation}
   On account of the transformation eq.(\ref{e6}) and the Heisenberg equation
eq.(\ref{e12}) , we obtain:
\begin{equation}
\label{e13}
\partial_t \Phi_i(x,t)-vC_i \partial_x \Phi_i(x,t)=
              i\sum_{j=1}^{4}[v(C_i-C_j)\theta_{ij}-2gC_{ij}]n_j(x,t)\psi_i(x,t)
             \exp[-i\sum_{k=1}^{4}\theta_{jk}\phi_k (x)]
\end{equation}
     By choosing
\begin{equation}
\label{e14}
\theta_{ij} =\frac{2g}{v} \frac{C_{ij}}{C_i-C_j} \;\;\ (C_i\neq C_j)
\end{equation}
\begin{equation}
\label{e15}
\theta_{ii}=0
\end{equation}
then $\Phi_i(x,t)$ satisfy the free-field equation .
 The Hamiltonian becomes diagonalized (here we suppose: $C_i\neq C_j$; if $C_i=C_j$,
 the Hamiltonian can be diagonalized only when $C_{ij}=0$):
\begin{equation}
\label{e16}
H'=iv\sum_{i=1}^{4} C_i \int \Phi_i^+(x)\partial_x \Phi_i (x) dx
\end{equation}
   The direct calculation shows that $\Phi_i (x)$ and $H'$ also satisfy
the Heisenberg equation
$i\partial_t {\Phi}_i(x,t)=[\Phi_i(x,t),H']$, so $\Phi_i(x,t)$ are really dynamic
 variables regarding $H'$.

  Therefore, by using the local $U(1)$ transformation  eq.(\ref{e6}),
the original Hamiltonian eq.(\ref{e1}) constructed by $\psi_i(x)$ with
anticommutation relations eq.(\ref{e2})-eq.(\ref{e4}) has been transformed
into the quadratic Hamiltonian eq.(\ref{e16}) in terms of the $\Phi_i(x)$
obeying q-deformed relations eq.(\ref{e7})-eq.(\ref{e9}). In the following,
we shall show how the method in[8,9,10] works to find the Bethe ansatz
wavefunction in a simple manner for $SO(5)$ massless Thirring model.

    Let us denote by $\mid n_1,n_2,n_3,n_4 >$ a eigenstate with
 $n_i$ $\Phi_i$-particles($i$=1,2,3,4),
it can be  expressed by
$$\mid n_1,n_2,n_3,n_4 >=\int ... \int \prod_{j=1}^{M} dx_j
\varphi(x_1,...,x_M)
\prod_{j_1=1}^{n_1}
  \Phi^+_1(x_{j_1})$$
\begin{equation}
\label{e17}
\times\prod_{j_2=1}^{n_2}\Phi^+_2(x_{M_1+j_2})\prod_{j_3=1}^{n_3}
\Phi^+_3(x_{M_2+j_3})
\prod_{j_4=1}^{n_4}\Phi^+_4(x_{M_3+j_4}) \mid 0 >
\end{equation}
where $M_i=n_1+n_2+...+n_i$ , $M=M_4$ and $\mid 0 >$ is the vacuum
defined by
\begin{equation}
\label{e18}
\psi_j(x) \mid 0 >=0
\end{equation}
or equivalently
\begin{equation}
\label{e19}
\Phi_j(x) \mid 0 >=0
\end{equation}
Substituting eq.(\ref{e17}) and eq.(\ref{e16}) into Schr$\ddot{o}$dinger
equation
\begin{equation}
\label{e20}
H' \mid n_1,n_2,n_3,n_4>=E_{n_1,n_2,n_3,n_4} \mid n_1,n_2,n_3,n_4>
\end{equation}
it yields equation for $\varphi(x_1,...,x_M) $
\begin{equation}
\label{21}
iv(\sum_{i=1}^{4}C_i \sum_{j_i=1}^{n_i}
\frac{\partial}{\partial {x_{M_{i-1}+j_i}}})
\varphi(x_1,...,x_M)=E_{n_1,n_2,n_3,n_4} \varphi(x_1,...,x_M)
\end{equation}
whose solution is:
$$\varphi(x_1,...,x_M)=A\exp(i\sum_{j=1}^{M}k_j x_j)$$
\begin{equation}
\label{e22}
E_{n_1,n_2,n_3,n_4}=-v(\sum_{i=1}^{4}C_i \sum_{j_i=1}^{n_i} k_{M_{i-1}+j_i})
\end{equation}
where $k_j$ and $A$ are constants. Since the constant $A$ is not essential,
we shall omit it hereafter. The Bethe ansatz wavefuntion $\hat{\varphi}(x_1,...,x_M)$
is defined by
$$\mid n_1,n_2,n_3,n_4>=\int ... \int \prod_{j=1}^{M} dx_j
\hat{\varphi}(x_1,...,x_M) \prod_{j_1=1}^{n_1}
  \psi^+_1(x_{j_1})$$
\begin{equation}
\label{23}
\times\prod_{j_2=1}^{n_2}\psi^+_2(x_{M_1+j_2})
  \prod_{j_3=1}^{n_3}\psi^+_3(x_{M_2+j_3})
  \prod_{j_4=1}^{n_4}\psi^+_4(x_{M_3+j_4}) \mid 0 >
\end{equation}
Substituting eq.(\ref{e6}) into eq.(\ref{e17}), by detail calculation,
we have
$$\mid n_1,n_2,n_3,n_4>=\int ... \int \prod_{j=1}^{M} dx_j
\varphi(x_1,...,x_M)
\prod_{1\le p<q\le 4} \prod_{j_p=1}^{n_p}
\prod_{j_q=1}^{n_q}\exp[i\theta_{pq}
 \theta(x_{M_{p-1}+j_p}-x_{M_{q-1}+j_q})]$$
$$\times\prod_{j_1=1}^{n_1}
\psi^+_1(x_{j_1}) \prod_{j_2=1}^{n_2}\psi^+_2(x_{M_1+j_2})
\prod_{j_3=1}^{n_3}\psi^+_3(x_{M_2+j_3})
\prod_{j_4=1}^{n_4}\psi^+_4(x_{M_3+j_4}) \mid 0 > $$
$$\sim  \int ... \int \prod_{j=1}^{M} dx_j \varphi(x_1,...,x_M)
 \prod_{1\le p<q\le 4} \prod_{j_p=1}^{n_p}\prod_{j_q=1}^{n_q} [1-itg\frac{\theta_{pq}}{2}
 \epsilon(x_{M_{p-1}+j_p}-x_{M_{q-1}+j_q})]$$
\begin{equation}
\label{e24}
\times\prod_{j_1=1}^{n_1}
  \psi^+_1(x_{j_1}) \prod_{j_2=1}^{n_2}\psi^+_2(x_{M_1+j_2})
  \prod_{j_3=1}^{n_3}\psi^+_3(x_{M_2+j_3})
  \prod_{j_4=1}^{n_4}\psi^+_4(x_{M_3+j_4}) \mid 0 >
\end{equation}
where $\theta(x)=0 \;\ (if\;\ x<0) ;
1 \;\ (if\;\ x>0)$ and
$\epsilon(x)=\theta(x)-\theta(-x)$ hereafter .
Thus, the Bethe ansatz wavefuntion $\hat{\varphi}(x_1,...,x_M)$  takes the form:
\begin{equation}
\label{e25}
\hat{\varphi}(x_1,...,x_M)=\exp[i\sum_{j=1}^{M}k_j x_j]
\prod_{1\le p<q\le 4} \prod_{j_p=1}^{n_p}\prod_{j_q=1}^{n_q} [1-itg\frac{\theta_{pq}}{2}
 \epsilon(x_{M_{p-1}+j_p}-x_{M_{q-1}+j_q})]
\end{equation}
which describes the many-body problem with $\delta$-interactions.

Suppose that $M$ particles move in a region with the length $L$.
For an arbitary $x_j\; (M_{p-1}\le j \le M_p)$.
Imposing the periodical boundary conditions (PBC), we have
\begin{equation}
\label{e26}
k_j L=-i\sum_{\stackrel{q\neq p}{q=1}}^{4}n_q
ln\frac{1-itg\theta_{pq}/2}{1+itg\theta_{pq}/2} +2l_j\pi \;\;\;\;\;\;\;\;\; (l_j\; integer)
\end{equation}
i.e.
\begin{equation}
\label{e27}
\label{bac}
k_j L=-\sum_{\stackrel{q\neq p}{q=1}}^{4}n_q\theta_{pq} +2l_j\pi \;\;\;\;\;\;\;\;\; (l_j\; integer)
\end{equation}
that is exactly the Bethe ansatz equation. Obviously, the local $U(1)$ transformation
eq.(\ref{e6}) much helps the derivation of the Bethe ansatz condition for the massless
Thirring model.

{\bf (III). Current realization of $Y(SO(5))$ }

    The $SO(5)$ algebra does have the current realization, however the
fermionic construction is not unique. In parallel to the diagonlization
of eq.(\ref{e1}) we shall show that the q-deformed operators $\Phi_i(x)$ shown
in eq.(\ref{e6}) also provides a realization of $SO(5)$ algebra, henceforth
Yangian associated with $SO(5)$.

    The original commutation relations of $Y(g)$ were given by Drinfled[17,18]
in the form:
\begin{equation}
\label{e28}
[I_\lambda,I_\mu]=c_{\lambda\mu\nu}I_\nu \;\;\;\;\;\;\;\;\\;\;\;\
[I_\lambda,J_\mu]=c_{\lambda\mu\nu}J_\nu
\end{equation}
\begin{equation}
\label{e29}
[J_\lambda,[J_\mu,I_\nu]]-[I_\lambda,[J_\mu,J_\nu]]
=h^2 a_{\lambda\mu\nu\alpha\beta\gamma} \{I_\alpha,I_\beta,I_\gamma\}
\end{equation}
\begin{equation}
\label{e30}
[[J_\lambda,J_\mu],[I_\sigma,J_\tau]]+[[J_\sigma,J_\tau],[I_\lambda,J_\mu]]
=h^2 (a_{\lambda\mu\nu\alpha\beta\gamma}c_{\sigma\tau\nu}
+a_{\sigma\tau\nu\alpha\beta\gamma}c_{\lambda\mu\nu})\{I_\alpha,I_\beta,I_\gamma\}
\end{equation}
where $c_{\lambda\mu\nu}$ are structure constants of a simple Lie algebra $g$,
$h$ is a constant and
\begin{equation}
\label{e31}
 a_{\lambda\mu\nu\alpha\beta\gamma}=\frac{1}{4!}c_{\lambda\alpha\sigma}
 c_{\mu\beta\tau}c_{\nu\gamma\rho}c_{\sigma\tau\rho} \;\;\;\;;\;\;\;\;
\{x_1,x_2,x_3\}=\sum_{i\not=j\not=k}x_{i}x_{j}x_{k}
\end{equation}
 For Lie algebra $SO(5)$, $Y(SO(5))$ is generated by
antisymmetric generators $\{I_{ab},J_{ab}\}$.
eq.(\ref{e28}) reads
\begin{equation}
\label{e32}
[I_{ab},I_{cd}]=i(\delta _{bc}I_{ad}+\delta _{ad}I_{bc}
-\delta _{ac}I_{bd}-\delta _{bd}I_{ac})
\end{equation}
\begin{equation}
\label{e33}
[I_{ab},J_{cd}]=i(\delta _{bc}J_{ad}+\delta _{ad}J_{bc}
-\delta _{ac}J_{bd}-\delta _{bd}J_{ac})
\end{equation}
$$
I_{ab}=-I_{ba}; \;\; J_{ab}=-J_{ba};\;\; (a,b,c,d=1,2,3,4,5)
$$

  Not all of the relations in eq.(\ref{e29})-eq.(\ref{e30}) are
  independent. After tedious calculation we can prove that there
  is only one independent ralation:
\begin{equation}
\label{e34}
[J_{23},J_{15}]=\frac{i}{24}h^2 (\{I_{13},I_{42},I_{45}\}+\{I_{12},I_{45},I_{34}\}
-\{I_{14},I_{42},I_{35}\}-\{I_{14},I_{34},I_{25}\})
\end{equation}
 where $J_{23}$ and $J_{15}$ are the Cartan subset.

All the other relations other than eq.(\ref{e28}) can be generated on the basis
of eq.(\ref{e34}) by using
Jacobi identities together with eq.(\ref{e32}) and eq.(\ref{e33}).
Therefore, for $Y(SO(5))$, eq.(\ref{e28})-eq.(\ref{e30}) can be expressed with
eq.(\ref{e32})-eq.(\ref{e34}) in such a simple manner. This conclusion can
also be verified by RTT relation independently through tremendious computation.

  The generators of $Y(SO(5))$ can be realized by fermion current algebra
as follows:
$$
I_{ab}=\int I_{ab}(x)dx \;\;\;\;\;;\;\
I_{ab}(x)=-\frac{1}{2}\psi^+(x)\Gamma^{ab}\psi(x)
 \;\;\;\;\;\;\;\;\;\;\;\;\;\;\;\;\;\;\;\;\;\;\
$$
$$
J_{ab}=T_{ab}+UJ_{ab}^{0} \;\;\;\;;\;\;
T_{ab}=\int dx\psi^+(x)\Gamma^{ab}\partial_{x}\psi(x)
\;\;\;\;\;\;\;\;\;\;\;\;\;\;\;\;\;\;\;\;\;\;\;\;\;\;\
$$
\begin{equation}
\label{e35}
\;\;\;\;\ J_{ab}^{0}= \int \int dxdy
\epsilon(x-y) I_{ac}(x)I_{cb}(y) \;\;\;\;\;\;\;\;\;\;\;\;\;\;\;\;\
\end{equation}
where $\Gamma^{ab}=-i\Gamma^{a}\Gamma^{b}$, $\Gamma^a$ are 4$\times$4 Dirac
matrices, $U=\pm\frac{i}{2}h$ ($h$ being arbitrary constant) and
$\psi(x)$ satisfies anticommutation relations eq.(\ref{e2})-eq.(\ref{e4}).
It can be checked that the set $\{I_{ab},J_{ab}\}$
satisfies algebraic relations eq.(\ref{e32})-eq.(\ref{e34}) of $Y(SO(5))$.

As given by Ref[6] if $\psi(x)=[c_\sigma(x),d_\sigma(x)]^T$,
then local generators $I_{ab}(x)$
of Lie algebra $SO(5)$ are expressed in terms of spin
$\vec{S}(x)=\frac{1}{2}(c^+(x)\vec{\sigma}c(x)+d^+(x)\vec{\sigma}d(x))$,
charge $Q(x)=\frac{1}{2}(c^+(x)c(x)+d^+(x)d(x)-2)$ and $\vec{\pi}^+(x)=-\frac{1}{2}
c^+(x)\vec{\sigma}\sigma_2 d^+(x)$ with
\begin{equation}
\label{e36}
\ I_{ab}(x)=\left(
\begin{array}{ccccc}
0&&&&\\
\pi_1^+(x)+\pi_1(x)&0&&&\\
\pi_2^+(x)+\pi_2(x)&-S_3(x)&0&&\\
\pi_3^+(x)+\pi_3(x)&S_2(x)&-S_1(x)&0&\\
Q(x)&i(\pi_1(x)-\pi_1^+(x))&i(\pi_2(x)-\pi_2^+(x))&i(\pi_3(x)-\pi_3^+(x))&0\\
\end{array}   \right)
\end{equation}
where the value of matrix elements on the upper right triangle are
determined by antisymmetry, $I_{ab}(x)=-I_{ba}(x)$.

Under the local $U(1)$ transformation eq.(\ref{e6}), the four-component
fermion field operators $\psi(x)=[\psi_1(x),\psi_2(x), \psi_3(x),\psi_4(x)]^T$
is changed into q-deformed
operator $\Phi(x)=[\Phi_1(x),\Phi_2(x),\Phi_3(x),\Phi_4(x)]^T$.
The generators of $Y(SO(5))$ is constructed by q-deformed fermionic
current algebra as follows:
$$
\overline{I}_{ab}=\int\overline{I}_{ab}(x)dx \;\;\;\;\;;\;\
\overline{I}_{ab}(x)=-\frac{1}{2}\Phi^+(x)\Gamma^{ab}\Phi(x)
$$
$$
\overline{J}_{ab}=\overline{T}_{ab}+U\overline{J}_{ab}^{0}\;\;\;\;\;\;;\;\
\overline{T}_{ab}=\int dx\Phi^+(x)\Gamma^{ab}\partial_{x}\Phi(x)
$$
\begin{equation}
\label{e37}
\;\;\;\;\;\;\;\;\;\;\ \overline{J}_{ab}^{0}=\int\int dxdy
\epsilon(x-y)\overline{I}_{ac}(x)\overline{I}_{cb}(y) \;\;\;\;\;\;\;\;\;\;\;\;\;\;\;\;\;\;\;\;\;\;\;\;\;\;\;\;\;\;\;\;\;\;\;\;\;\;\
\end{equation}
Substituting eq.(\ref{e37}) into eq.(\ref{e32})-eq.(\ref{e34}), we can obtain
the constraint conditions:
\begin{equation}
\label{e38}
\theta_{im}-\theta_{jm}=
                     \theta_{in}-\theta_{jn} \;\;\ (mod\;\ 2\pi)
\end{equation}
$$ \;\;\;\;\;\;\  (i,j,m,n=1,2,3,4)$$
where eq.(\ref{e38}) sets the conditon making $Y(SO(5))$ constructed
 by q-deformed field operator $\Phi(x)$, This indicates that the
 current realization of $Y(SO(5))$ is not unique.
  Careful calculation shows that there are three
free parameters in $\theta_{ij}$ under conditions
eq.(\ref{e10}), eq.(\ref{e11}) and eq.(\ref{e38}). So there exists an
additional freedom in $Y(SO(5))$. Substituting
eq.(\ref{e14}) into eq.(\ref{e38}), we obtain
\begin{equation}
\label{e39}
\frac{C_{im}}{C_i-C_m}-\frac{C_{jm}}{C_j-C_m}=
\frac{C_{in}}{C_i-C_n}-\frac{C_{jn}}{C_j-C_n}\;\ (mod\;\ 2\pi)
\end{equation}
$$ \;\;\;\;\;\;\  (i,j,m,n=1,2,3,4)$$

   In another words, $H'$ is the Hamiltonian expressed by  $\Phi_i(x)$ ,
under the transformation eq(\ref{e6}), it becomes $H$ where two-body
interaction appears. So the physical meaning of $U(1)$ transformation
in $Y(SO(5))$ is connected with two body interaction in massless
Thirring model with $SO(5)$ symmetry.

From the above analysis we see that there is a local $U(1)$
gauge-invariance in the construction of the current algebra
realization for $Y(SO(5))$.

Under local $U(1)$ transformation eq.(\ref{e6}), eq.(\ref{e36}) is changed into
\begin{equation}
\label{e40}
\overline{I}_{ab}(x)=\left(
\begin{array}{ccccc}
0&&&&\\
\overline{\pi}_1^+(x)+\overline{\pi}_1(x) &0&&&\\
\overline{\pi}_2^+(x)+\overline{\pi}_2(x)&-\overline{S}_3(x)&0&&\\
\overline{\pi}_3^+(x)+\overline{\pi}_3(x)&\overline{S}_2(x)&-\overline{S}_1(x)&0&\\
\overline{Q}(x)&i(\overline{\pi}_1(x)-\overline{\pi}_1^+(x))
&i(\overline{\pi}_2(x)-\overline{\pi}_2^+(x))
&i(\overline{\pi}_3(x)-\overline{\pi}_3^+(x))&0\\
\end{array}   \right)
\end{equation}
where the value of matrix elements on the upper right triangle
 are determined by antisymmetry, $\overline{I}_{ab}(x)=-\overline{I}_{ba}(x)$,
and

\[
\left(
\begin{array}{c}
\overline{S}_{1}(x)\\
\overline{S}_{2}(x)\\
\overline{S}_{3}(x)\\
\end{array}  \right) =
\left(
\begin{array}{ccc}
\cos(\frac{\alpha+\beta}{2}\phi(x))
\cos(\frac{\alpha-\beta}{2}\phi(x))
&-\sin(\frac{\alpha+\beta}{2}\phi(x))
\cos(\frac{\alpha-\beta}{2}\phi(x))&0\\
\sin(\frac{\alpha+\beta}{2}\phi(x))
\cos(\frac{\alpha-\beta}{2}\phi(x))
&\cos(\frac{\alpha+\beta}{2}\phi(x))
\cos(\frac{\alpha-\beta}{2}\phi(x)) &0\\
0&0&1\\
\end{array}   \right)
\left(
\begin{array}{c}
S_1(x)\\
S_2(x)\\
S_3(x)\\
\end{array} \right) \]
\begin{equation}
\label{e41}
 + \left(
\begin{array}{ccc}
-\sin(\frac{\alpha+\beta}{2}\phi(x))
\sin(\frac{\alpha-\beta}{2}\phi(x))
&-\cos(\frac{\alpha+\beta}{2}\phi(x))
\sin(\frac{\alpha-\beta}{2}\phi(x))&0\\
\cos(\frac{\alpha+\beta}{2}\phi(x))
\sin(\frac{\alpha-\beta}{2}\phi(x))
&-\sin(\frac{\alpha+\beta}{2}\phi(x))
\sin(\frac{\alpha-\beta}{2}\phi(x)) &0\\
0&0&1\\
\end{array}   \right)
\left(
\begin{array}{c}
N_1(x)\\
N_2(x)\\
N_3(x)\\
\end{array}  \right)
\end{equation}

\begin{equation}
\label{e42}
\left(
\begin{array}{c}
\overline{\pi}_{1}^+(x)\\
\overline{\pi}_{2}^+(x)\\
\end{array}  \right) =\exp(i\frac{\nu\phi(x)}{2})
\left(
\begin{array}{ccc}
\cos(\frac{\alpha+\beta}{2}\phi(x))&-\sin(\frac{\alpha+\beta}{2}\phi(x))\\
\sin(\frac{\alpha+\beta}{2}\phi(x))&\cos(\frac{\alpha+\beta}{2}\phi(x))\\
\end{array}   \right)
\left(
\begin{array}{c}
\pi_1^+(x)\\
\pi_2^+(x)\\
\end{array}  \right)
\end{equation}

\begin{equation}
\label{e43}
\overline{\pi}_3^+(x)=\exp(i\frac{\nu\phi(x)}{2})
[\cos(\frac{\alpha-\beta}{2}\phi(x))\pi_3^+
+\frac{1}{2}\Delta^+(x)\sin(\frac{\alpha-\beta}{2}\phi(x))]
\end{equation}
where $\phi(x)=\sum_{i=1}^{4} \phi_i(x)$, $\theta_{43}=\beta$,
$\theta_{12}=\alpha$, $\theta_{13}-\theta_{42}=\nu$, SC order
paremeter $\Delta^+(x)=-ic^+(x) \sigma_2 d^+(x)$ and AF order
parameter $\vec{N}(x)=\frac{1}{2}(c^+(x) \vec{\sigma} c(x)
-d^+(x)\vec{\sigma} d(x))$.

   From the above analysis we see that under the $U(1)$
   transformation eq(\ref{e6}),
eq.(\ref{e37}) still obey Yangian algebra as eq.(\ref{e35})
does if $\theta_{ij}$ satisfy the conditions
eq.(\ref{e10}),eq.(\ref{e11}) and eq.(\ref{e38}).
This indicates that there is a local $U(1)$ gauge-invariance in
the construction of the current realization for $Y(SO(5))$.
It turns out that after the transformation, there are local
 phase factors
in the current realization  of $Y(SO(5))$
 (shown by eq.(\ref{e41}),eq.(\ref{e42}) and eq.(\ref{e43}),
  but eq.(\ref{e37}) still satisfy
 $Y(SO(5))$ Yangian algebraic relations. i.e., there is a local $U(1)$
  gauge-invariance in such a current realization of
  $Y(SO(5))$.

    We also find that the transformation eq.(\ref{e6})
     can be used to diagonalize
the massless Thirring model with $SO(5)$
symmetry that will help to
understand the physical meaning of the introduced local
 $U(1)$ symmetry.

 In another words, $H'$ is the Hamiltonian expressed by  $\Phi_i(x)$ ,
under the transformation eq.(\ref{e6}), it becomes $H$ where
 two body interaction  appears.
 So the physical
meaning of $U(1)$ transformation in the current realization
of $Y(SO(5))$ connected with two body interaction
in this physical models.
Applying the transformation we find the local
 $U(1)$ gauge-invariance
in $Y(SO(5))$ explicitly .

 Noting that there are some non-trivial phase
 factors in the generators of Yangian, but they still satisfy
 the commutation relations of $Y(SO(5))$, i.e., there is
    a local $U(1)$ gauge-invariance in $Y(SO(5))$.

{\bf (IV).Conclusion and Acknowledge}

  Using a local $U(1)$ transformation connecting the
 four-component fermionic field operator
$\psi_i(x) $ with q-deformed one $\Phi_i(x) $ ,
it is helpful to
 diagonalize massless
 Thirring model with $SO(5)$ symmetry.
 The Bethe ansatz wavefuntion
is obtained in a simple manner.
It turns out that the current realization of $Y(SO(5))$
is not unique and exist a local $U(1)$ gauge transformation.
This shows the existence of a local $U(1)$ symmetry
in the current  realization of $Y(SO(5))$.
 Correspondingly, the transformation leads to the local
 $U(1)$-gauge invariance for $Y(SO(5))$.
 The explicit forms of phase factors
 for $SO(5)$ has been shown.

     The authors would like to thank Dr.Jing-Ling Chen
 helpful discussion .
This work is in part supported by NSF of China.

\pagebreak


\begin{thebibliography}{99}

\bibitem{1} S.C.Zhang, Science 275, 1089(1997).
\bibitem{2} S.Meixner, W.Hanke, E.Demler, and S.C.Zhang, Phys.Rev.Lett.79,
            4902(1997).
\bibitem{3} R.Eder, W.Hanke, and S.C.Zhang, Phys.Rev.B57, 13 781(1998).
\bibitem{4} C.L.Henly, Phys.Rev.Lett.80, 3590(1998).
\bibitem{5} S.Rabello, H.Kohno, E.Demler, and S.C.Zhang, Phys.Rev.Lett.80,
            3586(1998).
\bibitem{6} D.J.Scalapinp, S.C.Zhang, and W.Hanke, Phys.Rev.B58, 443(1998).
\bibitem{7} D.G.Shelton, and D.S$\acute{e}$n$\acute{e}$chal, cond-mat/97110251
            (unpublished).
\bibitem{8} Y.Komori, M.Wadati, J.Phys.Soc.Jpn.65(3), 722-72(1996).
\bibitem{9} Y.Komori, M.Wadati, Phys.Lett A 218, 42-48(1996).
\bibitem{10} M.Wadati, Phys.Rev.Lett.60, 635(1988).
\bibitem{11} F.D.M.Haldane, Exact Jastrow-Gutzwiller resonating-valence-bond
             ground state of spin-$\frac{1}{2}$ antiferronmagnetic Heisenberg
             chain with $\frac{1}{r^2}$ ,Phys.Rev.Lett.60, 635(1988).
\bibitem{12} M.L.Ge, K.Xue and Y.M.Cho,  Realization of Yangian in Quantum
             Mechanics and Application NIM-TP-97-12 to appear in Phys.Lett.A.
             Mo-Lin Ge and Kan Xue,Y(su(3)) in Quantum Mechimics ,
             Nankai Math. Inst Preprint.
\bibitem(13) V.D.Korepin, N.M.Bogoliubov and A.G.Izergin, Quantum Inverse
             Scattering Method and Correlation Funtion
\bibitem{14} F.D.M.Haldane, Yangian symmetry of integrable quantum chain
             with long-range interactions and a new describtion of states
             in conformal field theory,Phys.Rev.Lett.69, 2021(1992).
\bibitem{15} Zamolocdchikow A and Zamolodchikow Al, Ann.Phys.120 25(1979).
\bibitem{16} L.D.Faddeev, Sov.Sci.Rev.C1, 107(1980).
\bibitem{17} V.Drinfled, Sov.Math.Dokl.32, 254(1985).
\bibitem{18} V.Drinfled, Sov.Math.Dokl.36, 212(1985).
\end{thebibliography}
\end{document}